\documentclass[twocolumn,aps,showpacs]{revtex4}
\usepackage {psfig}
\usepackage{bm}

\begin{document}
\title{Three-dimensional rotation of even-even triaxial nuclei}

\author{Makito Oi} 
\email{m.oi@surrey.ac.uk}
\author{Philip M. Walker}
\affiliation{Department of Physics, University of Surrey, 
Guildford, Surrey GU2 7XH, United Kingdom}

\begin{abstract}
With the self-consistent three-dimensional
 cranked Hartree-Fock-Bogoliubov (3d-cranked HFB)
method, various types of rotational motion near the yrast line
 are investigated in an even-even nucleus 
in the $A\simeq 130$ mass region ($^{134}_{ 58}$Ce$_{76}$). 
The possibilities of chiral rotations, tilted-rotations, and dynamical
aspects of these rotations are discussed through the analysis of the
3d-cranked HFB solutions. 
Although a stable planar solution of the chiral rotation is
obtained, an aplanar chiral configuration is found to be unstable
when triaxial deformation is treated self-consistently.
\end{abstract}
\pacs{21.10.Re, 27.60.+j}
\maketitle

\addvspace{7mm}

Nuclei in the $A\simeq 130$ region are now of special interest
because it is suggested that a new form of nuclear rotation can be
expected: a chiral rotation \cite{FM97,DFD00}. 
This three-dimensional rotation
was predicted at first in odd-odd systems,
such as $^{134}_{59}$Pr$_{75}$, whose total angular momentum 
is built by a particle and a hole in valence orbitals in addition to a
triaxial rotor, 
under an assumption of the irrotational-flow moment of inertia. 
Their angular momentum vectors point along the shortest ($s$), 
longest ($l$) and  intermediate ($i$) axes of the triaxially deformed nucleus 
($\gamma\simeq 30^{\circ}$) 
\footnote{The Hill-Wheeler parameterization \cite{RS80} is employed for the
quadrupole deformation parameters, $\beta$ and 
$\gamma$, where the sign of $\gamma$ is opposite
 to the Lund convention.}, 
respectively. 
When the collective angular momentum is absent,
the total angular momentum consists only of the single-particle spins.
The corresponding state is called ``planar''
because the total angular momentum is in the $s-l$ plane.
The planar states correspond to ``bandheads'' of the chiral rotational
bands. Rotational members built upon the planar state
 are classified as ``aplanar'' states because these states have 
the genuine chiral configuration.
Possible experimental evidence of the nuclear chirality was reported
in $N=75$ odd-odd isotones in the same mass region \cite{SKC01},
through findings of pairs of near-degenerate $\Delta I = 1$ bands (so-called
``chiral doublets'').

It is an interesting question whether or not even-even nuclei 
can have the chiral configuration after breaking (Cooper) pairs.
To realise the nuclear chirality in even-even systems,
at least four quasi-particles must be excited; 
two  neutrons  and two protons. 
This multi-quasiparticle excited state is expected at high spin 
where rotation-alignment occurs due to the Coriolis force.
It is likely that such a ``simultaneous'' aligned state would be seen 
at higher spin, compared with the odd-odd chiral systems.
Experimentally, a candidate for chiral doublets in even-even nuclei
is suggested in $^{136}_{60}$Nd$_{76}$ \cite{MP02}, 
but there is as yet no detailed theoretical analysis about the possibility of 
chiral solutions in even-even nuclei, except a brief conference report
\cite{Fr02}.

The original model of the nuclear chirality is based on the
particle(hole)-plus-rotor model which admits 
the presence of a macroscopic rotor with solid triaxial deformation 
($\gamma\simeq 30^{\circ}$).
Although there is no concept of a rotor in fully microscopic models like
the HFB approach, the rotor spin can be treated as 
the collective part of angular momentum in the cranking model,
which can handle the collective and single-particle angular momentum
on the same footing.
An analysis 
of the chiral configuration in odd-odd systems was carried
out through the Tilted-Axis Cranking model (TAC) \cite{DFD00}.
Although each microscopic calculation (TAC)
is made for given parameters of shape and pairing gaps
as well as rotational frequency,
these parameters are carefully determined by
a macroscopic + microscopic approach (the Strutinski method)
in terms of the energy minimisation.

If nuclear chirality is produced in even-even systems by four
or more quasiparticle excitations, where the quasiparticles occupy orbitals
with very different orientation, it is expected that
the collective and individual (quasi-)particle motions 
would be strongly coupled through the Coriolis force.
Besides, the degree of triaxial deformation can be susceptible to 
the influence coming from rotation-alignments at high spin, especially
 in those nuclei whose Fermi level is in the upper shell (i.e.,
high-$\Omega$ Nilsson orbits) \cite{HMR79,OWA01}.
Therefore, if we attempt to describe the above physical situations,
purely microscopic and self-consistent treatments 
should be made with respect to nuclear shape and pairing.
In other words, 
for the aim of demonstrating the chirality in even-even systems,
it is necessary to show the emergence of a genuine three-dimensional
rotation together with the presence of substantial triaxial deformation.
Furthermore, compared to odd-odd systems, it is numerically advantageous 
for us to study even-even systems 
because the treatment of pairing is more straightforward.

In this paper, we report a fully microscopic and self-consistent study of 
three-dimensional nuclear rotations in even-even nuclei 
in the $A\simeq 130$ mass region near the yrast line \footnote{States with
highest spin for given energy.},
by means of the three-dimensional cranked
Hartree-Fock-Bogoliubov method (3d-cranked HFB) \cite{HO96,KO81}.

First of all, let us review the method.
The HFB equation is self-consistently solved 
with the pairing-plus-Q$\cdot$Q force
under constraints on particle numbers (proton and neutron), and  angular
momentum  (length and orientation). There are six  conditions
relevant to the latter constraints. They are,
$\langle\hat{J}_1\rangle=J\cos\theta,\langle\hat{J}_2\rangle=J\sin\theta\sin\phi,\langle\hat{J}_3\rangle=J\sin\theta\cos\phi$,
and $\langle B_{i}\rangle =0$ ($i=1,2,3$), with $B_{i}\equiv
\frac{1}{2}\left(\hat{Q}_{jk}+\hat{Q}_{kj}\right)$ ($i,j,k$: cyclic and
each index runs from 1 to 3).
The last three conditions are necessary when self-consistent calculations
are performed in shape, so as to fix the intrinsic coordinate
system set along the principal axes of the quadrupole tensor at rest
(i.e., $J=0$) \cite{KO81}.
Note that azimuthal ($\phi$) and polar ($\theta$) angles are measured, 
respectively, from the 1-3 plane and the 1-axis.
The nucleus $^{134}_{ 58}$Ce$_{76}$ is chosen for this study
because of the following three reasons: (i) it is in the $A\simeq 130$
region where chiral doublets are seen, (ii) it is known to be 
$\gamma$-soft in its ground state \cite{OG99},
and (iii) it has its Fermi levels for protons and neutrons located in
the high-$j$ intruder orbitals (h$_{11/2}$) with 
low- and high-$\Omega$ components, respectively.
According to  Baranger and Kumar \cite{BK65},
two major shells of the spherical Nilsson basis 
are needed for each sector of isospin (N=4 and 5, for proton and
neutron), in addition to  intruder orbits (i$_{13/2}$ both for proton
and neutron).
The force strengths for the quadrupole and pairing parts 
are determined through the (no-cranking) Nilsson + BCS model with the 
deformation parameters $(\beta,\gamma)=(0.162,0.0^{\circ})$
and gap energies $(\Delta_{p},\Delta_{n})=(1.069,0.881)$ [MeV]
\cite{MN}. The corresponding wave function is 
used as an initial state for the self-consistent calculations at $J=0$.
The HFB equation for an arbitrary spin ($J$) 
is consecutively solved by means of the method of steepest descent
\cite{On86} 
through the principal-axis cranking procedure $(\theta=0^{\circ})$
with a step $\Delta J=0.1\hbar$ until high spin
where a four quasi-particle state appears
with significant triaxial deformation.
The pairing gaps and quadrupole deformation parameters are
calculated in a fully self-consistent manner. For instance, they are
calculated to be  
$(\beta,\gamma)=(0.132, 25^{\circ})$ and 
$(\Delta_p,\Delta_n)=(0.00, 0.18)$ [MeV] at $J=30\hbar$. 
The self-consistent tilted-axis cranking calculations for $J$ start 
with this principal-axis cranked  HFB state as the initial state.
We confirmed that, similarly to the principal-axis cranking, the steepest 
descent method may be employed to solve the HFB equation
for given $\phi$ along the direction of $\theta$ 
from $0^{\circ}$ to $90^{\circ}$ with a step $\Delta\theta=0.5^{\circ}$.
$\phi$ is given every $1^{\circ}$
from $0^{\circ}$ to $90^{\circ}$.
In this way, physical quantities, such as the energy surface $E^J(\phi\theta)$,
quadrupole deformations $\beta^J(\phi\theta)$,
$\gamma^J(\phi\theta)$ and  pairing gaps $\Delta_{\tau}^J(\phi\theta)$
 ($\tau = \rm p,n$),
are calculated for the given spin $J$ as a function of tilt angles $(\theta,\phi)$.

\begin{figure}[htbp]
\begin{tabular}{c}
 \psfig{figure=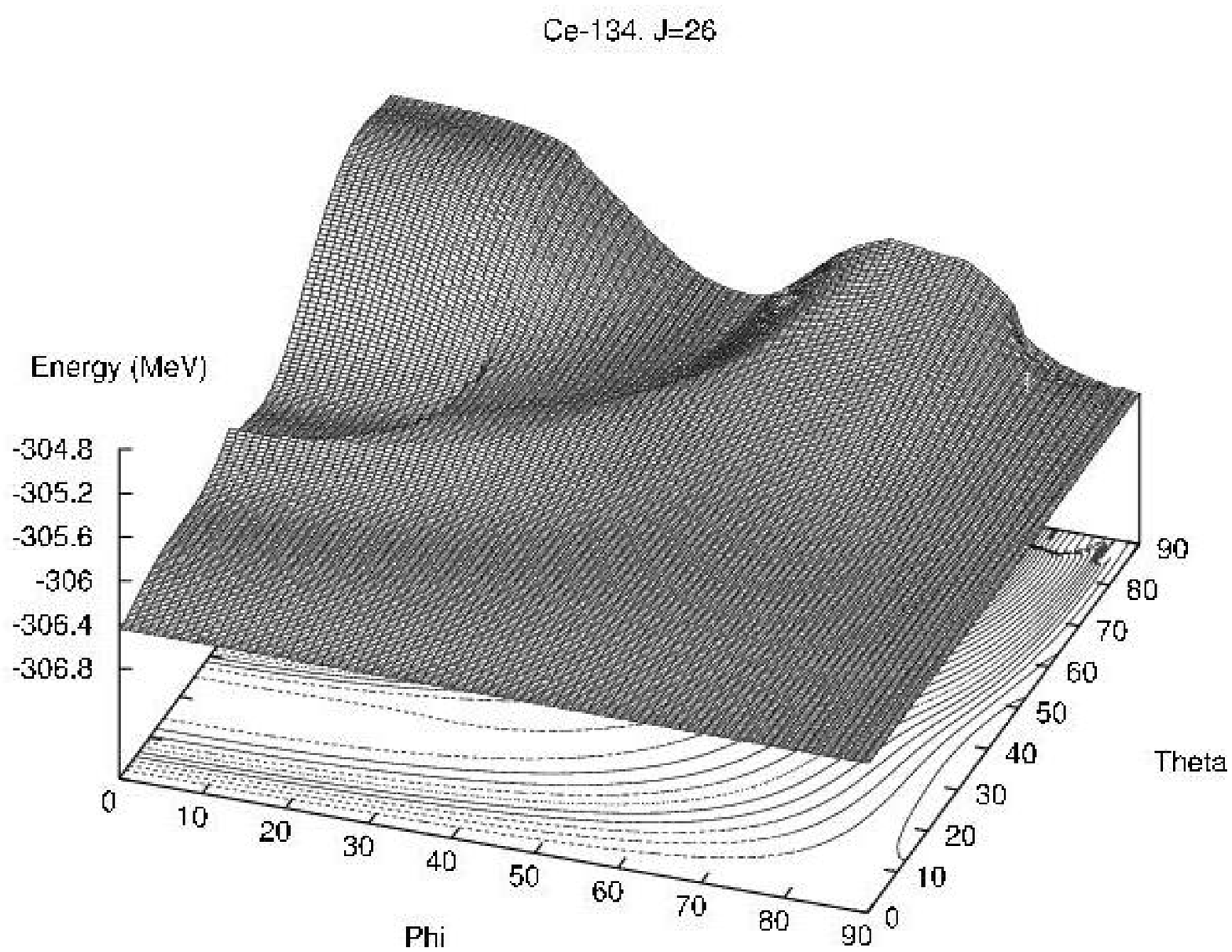,width=\textwidth}
 \\
 \psfig{figure=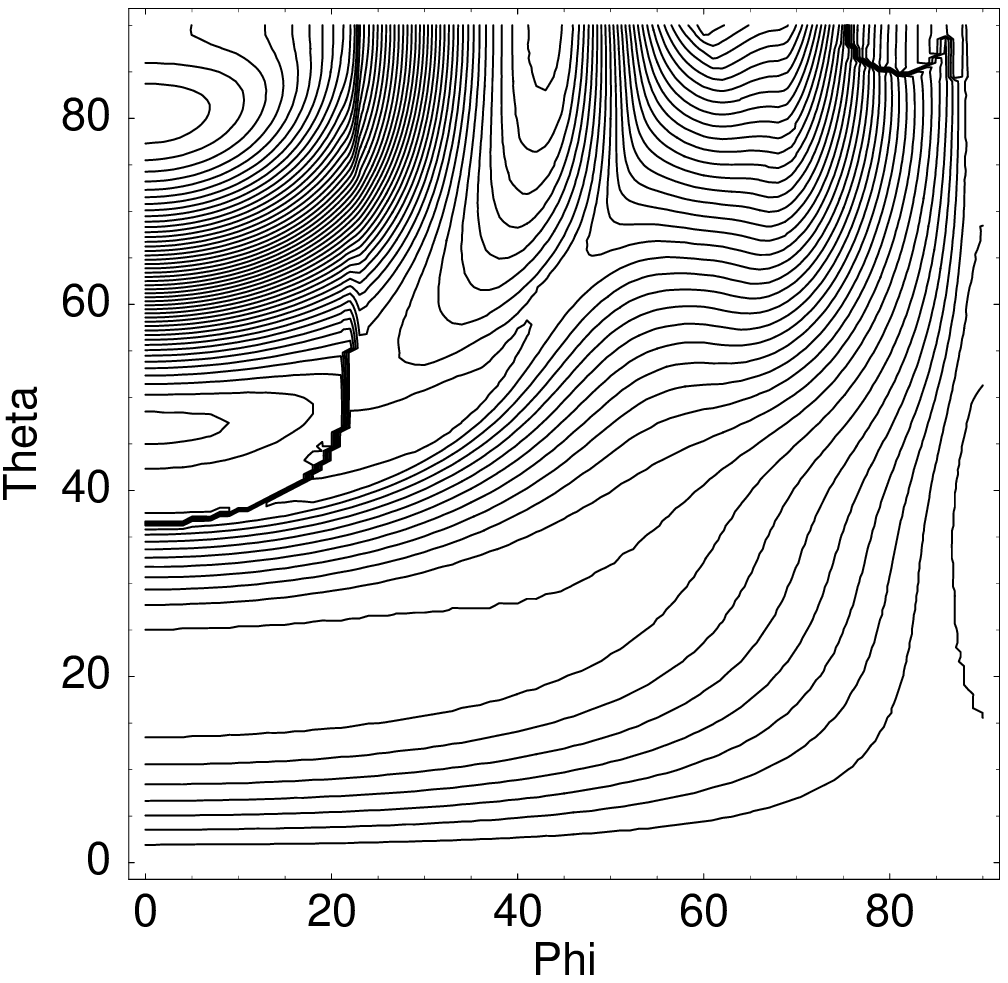,height=0.4\textheight}
\end{tabular}
 \caption{\label{fig0}Energy surface at $J=26\hbar$. The contour interval in the
lower panel is 50 keV. Discontinuities in the energy surface are
due to configuration changes, or level crossings.}
\end{figure}

At first, let us see the energy surface at $J=26\hbar$
(Fig.\ref{fig0}). 
There is a minimum at $(\theta,\phi)\simeq (90^{\circ},45^{\circ})$.
Deformations are calculated to be $(\beta,\gamma)=(0.16, 17^{\circ})$.
Major components of the single-particle spins 
come from both proton and neutron h$_{11/2}$ orbitals, and they are
respectively calculated as $(j_1,j_2,j_3)=(0.0,9.3,2.0)$ 
and (0.2,2.5,9.6).
(The unit for angular momentum is $\hbar$ here and hereafter.)
The 1-, 2-, and 3-axes correspond to the 
$i-$, $s-$, and $l-$axes, respectively,
in accordance with the $\gamma$ value.
The proton and neutron h$_{11/2}$ orbitals thus
point along the $s-$ and $l$-axes, respectively.
Collective rotation around the $i$-axis turns out not to be present.
This situation implies that the corresponding states
are the ``planar'' solutions of the chiral configuration.
(There is another minimum at $(\theta,\phi)\simeq
(47^{\circ},0^{\circ})$, but this state is identical to the minimum
we have just discussed above, from the analysis of the single-particle
configuration, shape parameters, etc.)

\begin{figure}[htbp]
\begin{tabular}{c}
 \psfig{figure=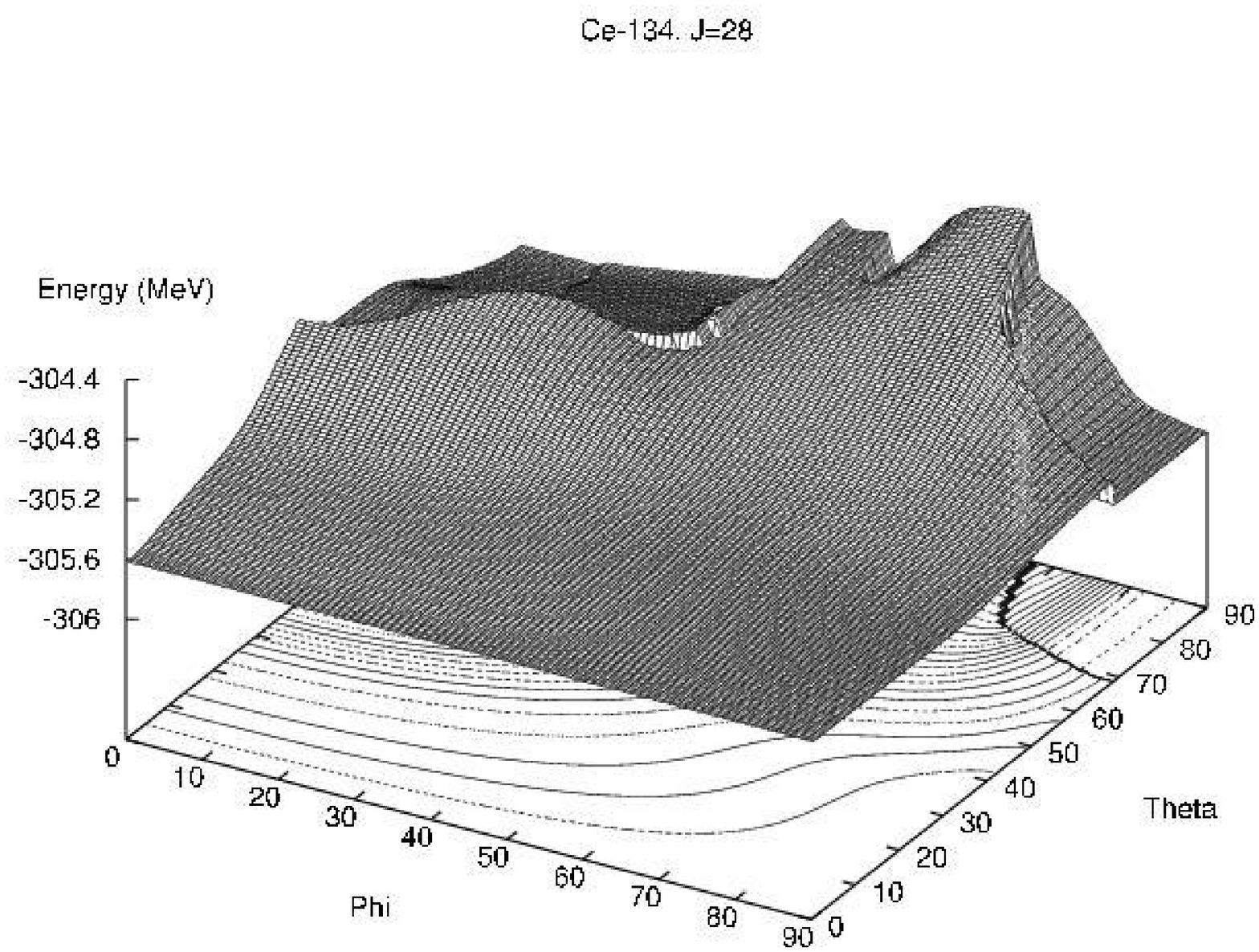,width=\textwidth}
 \\
 \psfig{figure=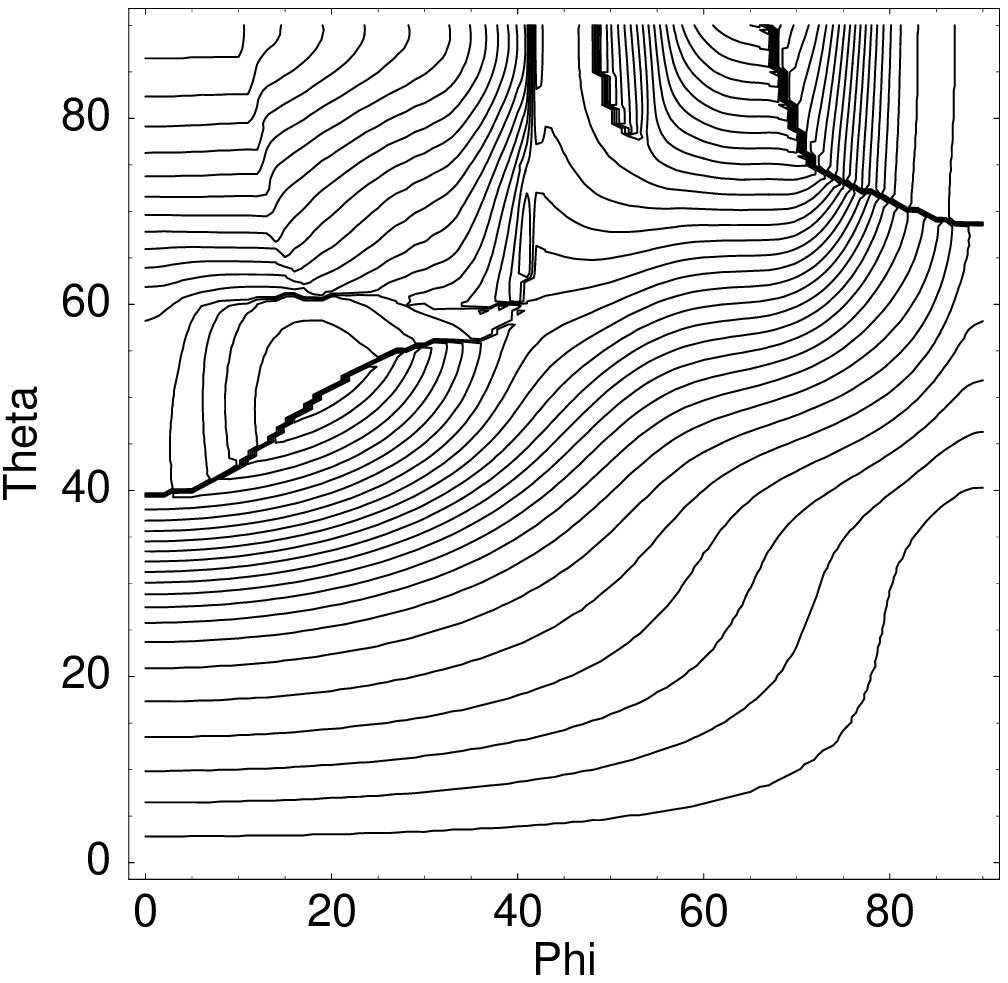,height=0.4\textheight}
\end{tabular}
 \caption{\label{fig2}Energy surface at $J=28\hbar$. The contour interval in the
lower panel is 40 keV. }
\end{figure}

The  ``aplanar'' chiral configuration 
 would be expected at $J=28\hbar$ if collective motion
could be added around the $i$-axis to the planar states in an ``adiabatic''
 manner. Adiabaticity here means the same  as in the context of the
discussion of rotational bands such as the ground state rotational bands
(g-bands). 
Analyses of the shape parameters, 
and of single-particle angular momenta for h$_{11/2}$ protons and
 neutrons over all the energy surface
imply that the chiral configuration \footnote{Note that a ``chiral
 configuration'' does not necessarily mean a ``chiral minimum'' or
 ``chiral solution''.} is realised
for $(\theta,\phi)\simeq (70^{\circ},45^{\circ})$ at $J=28\hbar$.
The corresponding quadrupole deformations are 
$(\beta,\gamma)=(0.13,32^{\circ})$,
and the alignments for proton and neutron h$_{11/2}$ orbitals are
(1.9, 9.3, 2.4) and (3.7, 4.3, 11.5), respectively.
However, as seen in Fig.\ref{fig2},
this state is  near a saddle point, implying that
this chiral configuration  is unstable.
The planar configuration, that forms an energy minimum at $J=26\hbar$,
 is quite susceptible to 
the collective rotation, through the  Coriolis force.

\begin{figure}[htbp]
\begin{tabular}{c}
 \psfig{figure=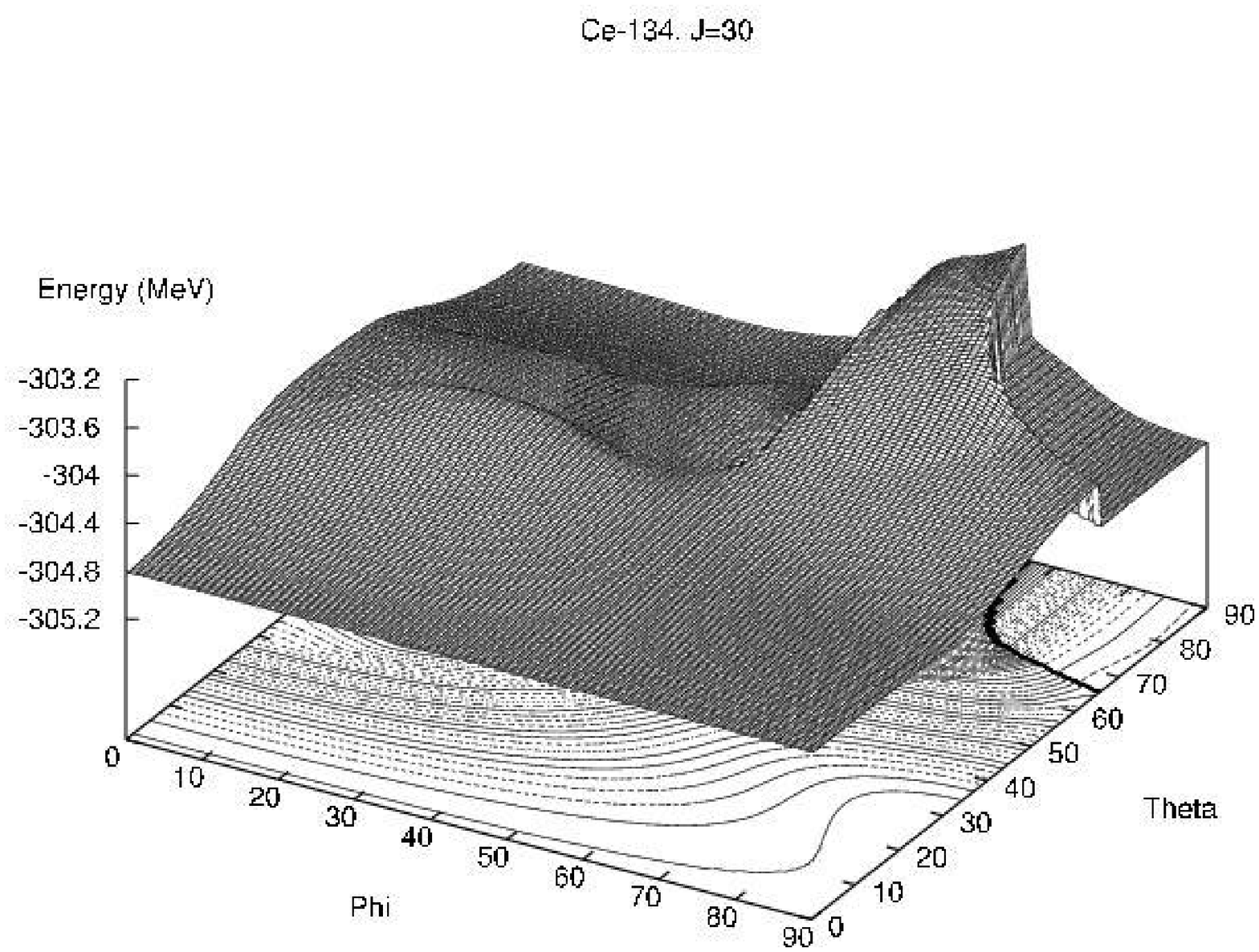,width=\textwidth}
 \\
 \psfig{figure=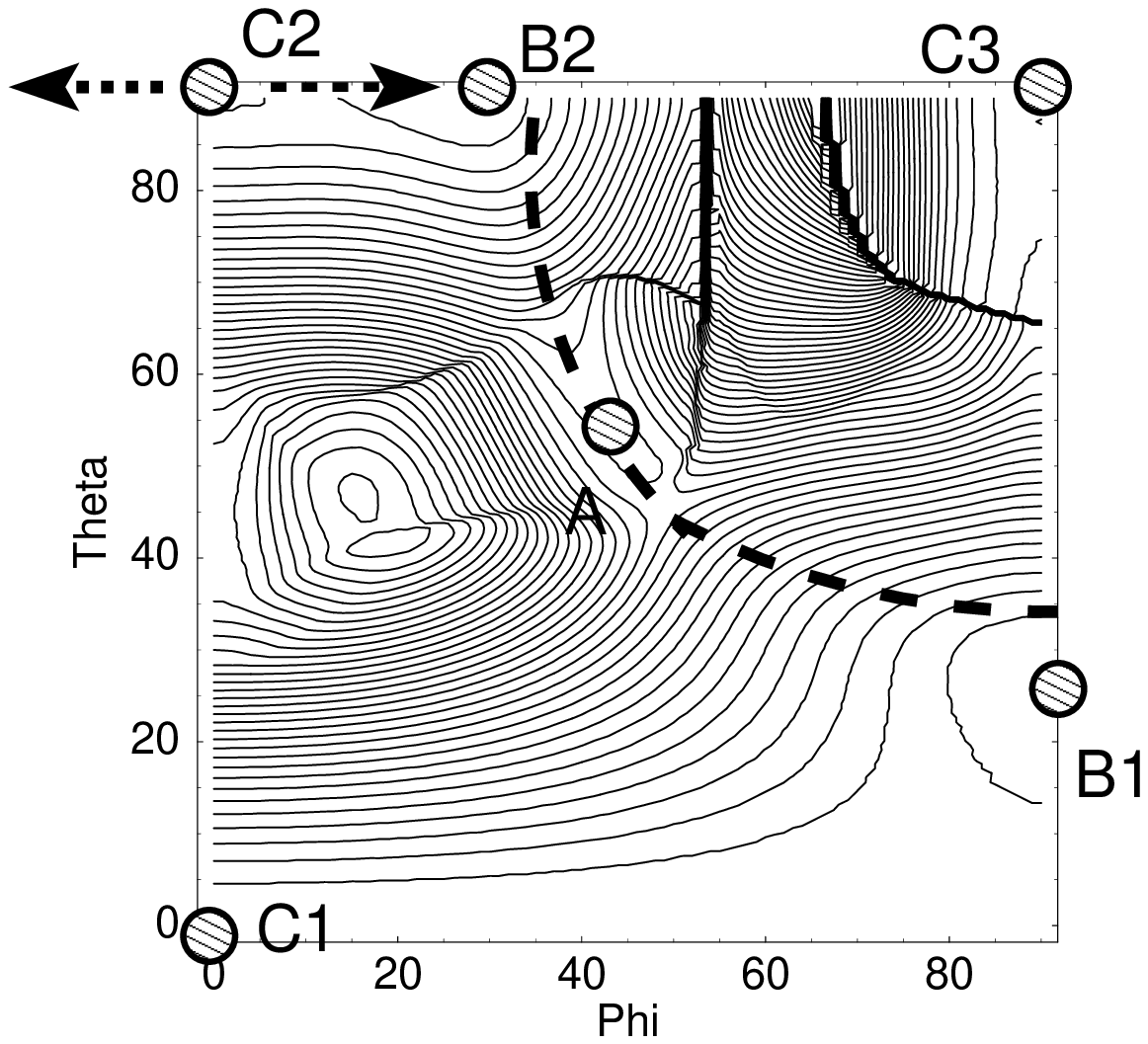,height=0.4\textheight}
\end{tabular}
 \caption{\label{fig1}Energy surface at $J=30\hbar$. 
The contour interval in the lower panel is 20 keV.
The trajectory of the zero mode of the oblate state ${\cal A}$ is 
schematically displayed by the dashed curve without an arrow.
The straight line with arrows indicates the nutation.}
\end{figure}

It is interesting to see what happens
when the angular momentum is further increased to $J=30\hbar$.
Fig.\ref{fig1} shows the energy surface at $J=30\hbar$. 
We have confirmed that 
similar structures to this energy surface 
are seen up to $J\simeq 38\hbar$.
There are six shallow minima found in the surface at $J=30\hbar$, including 
an extremely shallow one. 
Let us label these minima as follows: 
${\cal A}$: $(\theta,\phi)\simeq (58^{\circ},42^{\circ})$, 
${\cal B}_1$: $(\theta,\phi)\simeq (28^{\circ},90^{\circ})$, 
${\cal B}_2$: $(\theta,\phi)\simeq (90^{\circ},25^{\circ})$, 
${\cal C}_1$: $\theta\simeq 0^{\circ}$, 
${\cal C}_2$: $(\theta,\phi)\simeq (90^{\circ},0^{\circ})$, 
${\cal C}_3$: $(\theta,\phi)\simeq (90^{\circ},90^{\circ})$.
The groups ${\cal A,B}$ and ${\cal C}$ represent a three-dimensional rotating solution
(which we call ``3d-rotation''),
two-dimensional solutions (``tilted axis rotation'', or ``TAR''),
and one-dimensional solutions (``principal axis rotation'',
or ``PAR''), respectively.
Let us discuss these minima one by one.

{\it The minima ${\cal C}_i \ (i=1,2,3)$}:
This group is classified as one-dimensional rotation, or PAR.
The minima are located at
$(\theta,\phi)\simeq (0^{\circ},-), (90^{\circ},0^{\circ})$, and
$(90^{\circ},90^{\circ})$ for ${\cal C}_1$, ${\cal C}_2$, and ${\cal C}_3$, respectively.
The $\beta$ deformation values for these minima are almost equal
($\beta\simeq 0.13$).
The dominant components of total angular momentum come from
proton and neutron h$_{11/2}$ orbitals in either case.
The $\gamma$ deformation values are about $25^{\circ}, 30^{\circ},$ and
$-31^{\circ}$
for ${\cal C}_1 $, ${\cal C}_2$, and ${\cal C}_3$, respectively, 
so that all of them are strongly triaxially deformed. 
${\cal C}_1$ and ${\cal C}_2$ are identical from the analyses of 
their single-particle configurations, energy, deformation, etc.
They show a rotation around the $i$-axis while 
${\cal C}_3$  shows a rotation around the $s$-axis.
These two types of one-dimensional rotation can be classified
by  analogy with classical mechanics.
The moment of inertia for a rigid body (${\cal J}^{\text{rig}}$) and  
irrotational flow (${\cal J}^{\text{irr}}$) are then given as
${\cal J}^{\text{rig}}_{n}\propto 1-\sqrt{\frac{5}{4\pi}}\beta\cos(\gamma-\frac{2\pi}{3}n)$ and ${\cal J}^{\text{irr}}_n \propto \beta^2\sin ^2 (\gamma - \frac{2\pi}{3}n)$, respectively ($n=1,2,3)$. At $\gamma=30^{\circ}$,
the $i$-component of the moment of inertia becomes largest for
the irrotational flow, while it is the $s$-component for the rigid body.
${\cal C}_1$ is thus like ``irrotational flow''
and ${\cal C}_3$ is like ``rigid body'' rotation.
These two types are energetically degenerate in the present case, so
that the irrotational and rigid-body characters coexist
in this high-spin state.

{\it The minima ${\cal B}_i \ (i=1,2)$}:
The attribute of this group is two-dimensional rotation, or TAR.
${\cal B}_1$ at $(\theta,\phi)\simeq (28^{\circ},90^{\circ})$ and 
${\cal B}_2$ at $(90^{\circ},25^{\circ})$ are likely to be the
same states (that is, just a rearrangement of the principal axes).
Their deformations are calculated to be 
$(\beta,\gamma)\simeq(0.13,29^{\circ})$.
The 1-, 2- and 3-axes correspond to $i$-, $s-$ and $l$-axes, respectively.
Again proton and neutron h$_{11/2}$ orbitals play a leading role in building
total angular momentum. Their components are (6.2,7.0,0.0) for protons
and (11.2,3.8,0.0) for neutrons.
The neutron h$_{11/2}$ orbits have substantial components along the $i$-axis,
making
the total angular momentum vector for ${\cal B}_i$ lie in the $s-i$ plane.
This configuration is different from the ``planar'' chiral configuration 
having the total angular momentum vector in the $s-l$ plane.
The energy surface around $\cal B$ is shallow along one direction,
for example, along $\phi$ for ${\cal B}_2$. This structure implies
the possibility of fluctuation around the TAR.

{\it The minimum  ${\cal A}$,}
$(\theta,\phi)\simeq (58^{\circ},42^{\circ})$:
From the tilt angles, this ``minimum'' {\it seems} to correspond to the
state with 3d-rotation, but in fact this is not a 
3d-rotating state.
The quadrupole deformations are calculated to be
$(\beta,\gamma)=(0.12,60^{\circ})$, so that the corresponding nuclear
shape is oblate, that is, axially symmetric rather than triaxial.
The major contribution to the total angular
momentum comes from the proton and neutron h$_{11/2}$ orbits.
They are $(4.8, 8.3, 1.4)$ for protons, and
$(10.1, 4.0, 6.7)$ for neutrons. 
In this case, the shortest axis (i.e., the symmetry axis) corresponds to the
2-axis. 
The alignment along the $s$-axis comes predominantly from the proton
h$_{11/2}$ component, which is consistent with the chiral configuration
in odd-odd systems. However, the neutron h$_{11/2}$ component
contributes to building angular momentum  along
both the 1- and 3-axes.
This is because the principal axes of the quadrupole tensor 
along both of the coordinates are  identical, due to
the axial symmetry around the $s$-axis.
Therefore, this state is not  like the chiral rotation, but
rather corresponds to TAR of an oblate deformed nucleus.
This minimum ${\cal A}$ is extremely shallow (depth is of the order of 10 keV),
so that the adjacent states are energetically degenerate,
implying the presence of a ``zero mode''. Such a dynamical motion
will be discussed later.

The group ${\cal B}$ contains the lowest minima (hence the possible ``yrast'' state)
and the group ${\cal C}$ has the second lowest ($E_{\rm ex}\simeq$ 10 keV with
respect to the energy of ${\cal B}_i$). At the minimum ${\cal A}$, 
$E_{\rm ex} \simeq 300$ keV. 
As is easily seen from the energy surfaces being very shallow, it is more
natural to think that these states are coupled dynamically \footnote{There
are arguments that the large mass parameter can suppress the dynamical motions.
A brief discussion will be given later in this paper.}.
Namely, the state observed in reality will be a superposition of
these minima.
Nevertheless, it is useful and meaningful physically to classify
these dynamical states in terms of several simpler modes.
Let us now discuss the dynamical aspects of these minima.

First of all, a fluctuation around ${\cal B}_i$ should be studied
because they are the lowest minima.
The potential-well, for instance,  
around ${\cal B}_2$ extends one-dimensionally 
toward the direction of $\phi$,
indicating a swing motion around TAR.
This dynamical motion can be called ``nutation''.
The nutation is likely to occur
for $|\phi|\alt 40^{\circ}$ with $\theta=90^{\circ}$ in the case of
${\cal B}_2$. 
The signature symmetry, a symmetry with respect to $\pi$-rotation
around one of the principal axes, say the $i$-axis, is broken
in TAR. However, the nutation plays a role in restoring the symmetry
dynamically,  as seen in the multi-band crossings in the
$A\simeq 180$ region \cite{OAHO00}.
Another oscillatory motion is expected also with respect to $\theta$,
although this degree of freedom can be treated as  a minor correction
to the nutation. 
For these dynamical states, the
quadrupole deformations are calculated to be almost
constant and decoupled from the tilting degrees of freedom.

The second type of dynamical motion is expected around the
minimum $\cal A$ along the shallow plateau.
Because of the oblate deformation, all the states rotated
around the $s$-axis  are energetically degenerate. 
This sort of collective rotation
is regarded as a zero mode, which is closely connected
to the restoration of a continuous symmetry. In this case,
the axial symmetry broken by the TAR of an oblate deformed
state is restored by this zero mode.
Noting the convention of the tilt angles, 
i.e, $(J_1/J, J_3/J) = \left(\cos\theta, \sin\theta\cos\phi\right)$,
we rotate the state $\cal A$ around the $s$-axis 
(corresponding to the 2-axis in this case) by $\alpha$.
The rotated state is then expressed as
$\theta(\alpha)=\arccos (J_1^{\cal A}
\cos\alpha/J -\ J_3^{\cal A}\sin\alpha/J )$
and
$\phi(\alpha)=\arccos
\left\{
\csc\theta(\alpha)
(J_1^{\cal A}\sin\alpha/J + J_3^{\cal A}\cos\alpha/J)
\right\}$. 
Here, $(J_1^{\cal A}, J_2^{\cal A},J_3^{\cal A})$ 
denotes the components of total angular momentum for the minimum $\cal A$.
The trajectory $(\theta(\alpha),\phi(\alpha))$
is drawn  in Fig.\ref{fig1}, showing that
the shallow plateau near $\cal A$ corresponds to the
zero mode rotation.
Although the oblate deformed solution collapses beyond the plateau,
the structure of the energy surface between the minima ${\cal B}_i$
seems to retain the influence from this zero mode.
If the tunnelling between these minima through ${\cal A}$ has a
physical meaning, it should be related to the restoration of the 
broken symmetry, as previously discussed by Bonche et al. \cite{BDF91}.

These dynamical modes are related to restoration of symmetries.
In the first case, the nutation of the triaxial-deformed state restores
the discrete signature symmetry, while
the zero mode of the oblate-deformed TAR state restores
the continuous axial symmetry in the second.
The explicit treatment of these modes should be made by means of the
generator coordinate method, which it is planned to achieve in the future.

In summary,
the self-consistent 3d-cranked HFB equation is solved
for a $\gamma$-soft nucleus $^{134}$Ce, and
several solutions are obtained for one and two-dimensional
rotations (PAR, TAR, and planar). 
All the solutions are built upon four quasi-particle states
of h$_{11/2}$ orbitals for protons and neutrons. 
The chiral configuration is found as a planar state
(i.e., the ``bandhead'' of the chiral rotational band), 
but its aplanar partner turns out to be unstable in the present case.
As the angular momentum increases,
the corresponding configuration evolves into an oblate deformed state
whose rotational motion is TAR-like.
This oblate state is seen over a relatively 
wide range of angular momentum at high spin ($30\hbar \alt J \alt 38\hbar$).
As a general characteristic, 
it seems that  the
rotational motion  of $^{134}$Ce at high spin 
tends to be dynamical and non-uniform (nutation).
Furthermore, it should be noted that the chiral solutions in odd-odd
systems are also in quite shallow minima in the energy surface 
(barrier heights range from 50 to 100 keV) \cite{DFD00,Fr02}.
However, there is an argument in the framework of the particle-rotor model
that the tunnelling between chiral minima is suppressed when
the mass parameter is large \cite{Fr02}.
More studies may be necessary 
 to understand the dynamical aspects
of nuclear chirality, particularly
 in the context of the coupling between deformation
and orientation degrees of freedom,  in theory and hopefully in experiment.

M.O. thanks Professor H. Flocard for discussions that motivated
us to study nuclear chirality. 
He is also grateful to Professor P.-H. Heenen for fruitful discussions
and the information of Ref.\cite{BDF91}.
Our appreciation to Dr. P.H. Regan is acknowledged
for the information on Ref. \cite{MP02}.
This work is supported by an EPSRC advanced research fellowship
GR/R75557/01.

\end{document}